\documentclass[epj-spec]{svjour}
\usepackage{graphics}

\begin{document}

\title{Size dependence of multipolar plasmon resonance frequencies and
damping rates in simple metal spherical nanoparticles}
\subtitle{}
\author{A. Derkachova \and K. Kolwas}
\institute{Institute of Physics, Polish Academy of Sciences,\\
Al. Lotnik\'{o}w 32/46, 02-668 Warsaw, Poland}

\abstract{ Multipolar plasmon oscillation frequencies and
corresponding damping rates for nanospheres formed of the simplest
free-electron metals are studied. The possibility of controlling
plasmon features by choosing the size and dielectric properties of
the sphere surroundings is discussed. Optical properties of the
studied metals are described within the Drude-Sommerfeld model of
the dielectric function with effective parameters acounting for the
contribution of conduction electrons and of interband transitions.
No approximation is made in respect of the size of a particle;
plasmon size characteristics are described rigorously. The results
of our experiment on sodium nanodroplets \cite{DerkachovaKolwas} are
compared with the oscillation frequency size dependence of dipole
and quadrupole plasmon.  
} 

\maketitle

\section{Introduction}

The possibility of excitation and observation of surface plasmons in
spherical metal particles is a subject of continuously increasing
interest. It is connected with a wide range of applications of
plasmon excitations in nanotechnology, biophysics, biochemistry etc.
The most attractive feature of the surface plasmon resonances is the
concentration of electromagnetic field energy near the particle
surface. The Surface Enhanced Raman Spectroscopy is based on this
phenomenon \cite{KneipKneip,Moskovits}. SERS spectroscopic
techniques allow spectral studying of single molecules, particles
and cells \cite{Kneip,NieEmery}. Small noble metal spheres (with
size from 20 to 120 nanometers), introduced into an investigated
biomaterial, can be used as markers of some specific biomolecules,
tissue cancer changes or viruses \cite{ShultzSmith,ShultzMockSmith}.

The frequency dependence of the optical properties of a simple bulk
metal (alkali metal) change with free electron density, electron
relaxation rates, and the contribution of bound electrons to the
polarizability  \cite{KittelEng,KreibigVollmer}. The simplest model
for the dielectric function of bulk metal is the Drude-Sommerfeld
model of free electron gas. The optical properties of metal
nanospheres, as well as of others nanostructured metal materials,
are in addition geometry and (or) size dependent. These futures are
caused by the confinement of the electron gas resulting from the
presence of metal-dielectric boundary. In particular, optical
properties of spherical metal particles are characterized by size
dependent discrete eigenfrequencies. These eigenfrequencies can
manifest as resonances in the optical response of a sphere to the
external electromagnetic field. The complex eigenfrequencies define
the plasmon oscillation frequencies and the damping rates of
collective surface electron density oscillations which can be
excited by the external electromagnetic field. In contrast to the
flat metal surface, the curved surface enables the direct optical
excitation of surface plasmons.

In this paper we present a solution of the eigenproblem of
nanospheres formed of the simplest free-electron metals. The
analysis is concentrated on the influence of size and of material
parameters upon the multipolar plasmon features. Optical properties
of the studied metals are described within the Drude-Sommerfeld
model of the dielectric function with effective parameters
accounting for the contribution of interband transition to the
dielectric properties of metal. No approximation is made in respect
of the size of a particle; plasmon size characteristics are
described rigorously. We discuss the role of the material parameters
characterizing the electromagnetic properties of nanospheres in
controlling of plasmon features. We compare the expected size
dependence of plasmon oscillation frequency of dipole and quadrupole
plasmon with the results of our experiment on sodium nanodroplets
\cite{DerkachovaKolwas}.

\section{Eigenvalue problem for a metal sphere}

The eigenvalue problem is formulated in absence of external fields.
The eigenvalues result from the condition that the harmonic
solutions of Maxwell equations exist in both; the metal sphere and
its dielectric surroundings. The discrete complex frequencies of
electromagnetic fields result from the continuity relations (in
spherical coordinates) at the sphere boundary of the transverse
magnetic solutions of Maxwell equations (TM polarization). These
fields are coupled to the collective surface electron density
oscillations at the sphere surface, that are called surface
plasmons. The eigenfrequencies problem was presented in more detail
e.g. in \cite{Halevi} for the flat metal-dielectric interface and
e.g. in \cite{Halevi,KolwasDerkachova2} for the spherical interface.
At the flat boundary, the surface plasmon dispersion relation can be
obtained in a simple analytical form. The wave vector of a surface
plasmon wave $k_{sp}$ \cite{Halevi}:

\begin{equation}
k_{sp}=\frac{\omega }{c}\sqrt{\frac{\varepsilon _{m}(\omega )\varepsilon
_{d}(\omega )}{\varepsilon _{m}(\omega )+\varepsilon _{d}(\omega )}}
\label{flat}
\end{equation}

where $\varepsilon _{m}(\omega )$ and $\varepsilon _{d}(\omega )$
are the dielectric function of the metal and of the dielectric
surroundings respectively. For free-electron metal described by the
relaxation-free Drude dielectric function: $\varepsilon
_{m}(\omega)=1-\omega _{p}^{2}/\omega ^{2}$ , and $\varepsilon
_{d}(\omega )=1$, the dispersion relation \ref{flat} leads to the
well known "surface-plasmon frequency" at $\omega =$ $\omega
_{p}/\sqrt{2}$ \cite{Halevi}.

However, in the case of a spherical boundary, the plasmon dispersion
relation results from solution of the dispersion equation in complex form:%
\begin{equation}
\sqrt{\varepsilon _{in}}\xi _{l}^{\prime }\left( k_{out}R\right) \psi
_{l}\left( k_{in}R\right) -\sqrt{\varepsilon _{out}}\xi _{l}\left(
k_{out}R\right) \psi _{l}^{\prime }\left( k_{in}R\right) =0,
\label{DispEquation}
\end{equation}%
with $l=1,2,3...$where the wave numbers $k_{in}$, and $k_{out}$ are equal to:%

\begin{eqnarray}
k_{in} &=&\frac{\omega }{c}\sqrt{\varepsilon _{in}(\omega )}, \\
k_{out} &=&\frac{\omega }{c}\sqrt{\varepsilon _{out}(\omega )}.
\end{eqnarray}

$\varepsilon _{in}$ and $\varepsilon _{out}$ are dielectric
functions of the investigated metal sphere and of the dielectric
environment respectively, and define the corresponding refraction
coefficients: $n_{in}=\sqrt{\varepsilon _{in}}$ and
$n_{out}=\sqrt{\varepsilon _{out}}$. $\psi _{l}\left( z\right) $ and
$\xi _{l}\left( z\right) $ are Riccati-Bessel spherical functions
which can be expressed by the Bessel $J_{l+\frac{1}{2}}(z)$, Hankel
$H_{l+\frac{1}{ 2}}^{\left( 1\right) }(z)$ and Neuman
$N_{l+\frac{1}{2}}(z)$ cylindrical
functions of the half order, defined (e.g. in \cite{BornWolf}) as:%

\begin{eqnarray}
\psi _{l}(z) &=&z\cdot j_{l}(z)=z\sqrt{\frac{\pi }{2z}}J_{l+\frac{1}{2}}(z),
\label{RicBes1} \\
\xi _{l}(z) &=&\psi _{l}(z)-i\cdot \chi _{l}(z)=z\cdot h_{l}^{(1)}(z)=z\sqrt{%
\frac{\pi }{2z}}H_{l+\frac{1}{2}}^{(1)}(z),  \label{RicBes3} \\
\chi _{l}(z) &=&z\sqrt{\frac{\pi }{2z}}N_{l+\frac{1}{2}}(z).  \label{RicBes2}
\end{eqnarray}%

Solutions of the dispersion equations \ref{DispEquation} for each
$l$ mode exist only for the complex frequencies of the TM
(transverse magnetic) polarized electromagnetic field at the sphere
boundary, at $r=R$ \cite{Halevi,KolwasDerkachova2}:
\begin{equation}
\Omega _{l}(R)=\omega _{l}^{\prime }(R)+i\cdot \omega _{l}^{\prime \prime
}(R),  \label{ComplexFreq}
\end{equation}%
and can be found numerically for known dielectric functions of the
metal sphere $\varepsilon _{in}(\omega )$ and its dielectric
surroundings $\varepsilon _{out}(\omega )$. $\omega _{l}^{\prime
}(R)$ are the oscillation frequencies of TM electromagnetic field at
the surface in mode $l=1,2,3...$. $\omega _{l}^{\prime \prime }(R)$
are the damping frequencies of these oscillations, and are the
convolution of the radiative damping and of electron relaxation
processes.

\section{Drude-Sommerfeld model of the dielectric function}

Some of the metal properties, including the optical properties, can
be described within the simple free-electron gas Drude-Sommerfeld
model of the dielectric function. In the framework of this model,
with an external field applied, the conduction electrons move freely
between independent collisions occurring at the average rate of
$\gamma $. The frequency dependent dielectric function $\varepsilon
(\omega )$ predicted by Drude-Sommerfeld model:%

\begin{equation}
\varepsilon (\omega )=\varepsilon _{\infty }-\frac{\omega _{p}^{2}}{\omega
^{2}+i\gamma \omega },  \label{DrudeDielectricFunction}
\end{equation}%

includes the contribution of the bound electrons to the
polarizability by introducing phenomenological parameter
$\varepsilon _{\infty }$. This parameter equals $1$ only if the
conduction band electrons contribute to the dielectric
properties. The plasma frequency $\omega _{p}$ is given by:%

\begin{equation}
\omega _{p}=\sqrt{\frac{Ne^{2}}{\varepsilon _{0}m^{\ast }},}
\end{equation}%

where $N$ and $m^{\ast }$ are the density of conduction electrons
and the electron effective mass respectively.

In order to solve the dispersion equation \ref{DispEquation} with
respect to the frequency, we assumed that for the best free-electron
metal: $\varepsilon _{in}(\omega )=\varepsilon (\omega )$ (eq.
\ref{DrudeDielectricFunction}) with the following parameters:
$\varepsilon _{\infty }^{Na}=1.06$ \cite{Sievers}, $\omega
_{p}^{Na}=5.6$ $eV$ \cite{KittelEng} and $\gamma ^{Na}=0.03$ $eV$
for sodium, $\varepsilon _{\infty }^{Li}=5.843$, $\omega
_{p}^{Li}=8$ $eV$ \cite{KittelEng} and $\gamma ^{Li}=0.05$ $eV$ for
lithium, and $\varepsilon _{\infty }^{Cs}=1.8,$ $\omega
_{p}^{Cs}=3.4$ $eV$ \cite{KittelEng} and $\gamma ^{Cs}=0.03$ $eV$
for cesium. The dielectric function \ref{DrudeDielectricFunction}
for sodium (solid line in Fig. 1 a) and b)) reproduces the
optical constants $n$ measured for liquid and solid sodium%
\cite{InagakiArakawaBirkhoff,InagakiArakawaEmerson} (open and closed
circles on Fig. 1) quite well. However, the dielectric properties
for lithium \cite{InagakiArakawaEmerson} (squares in Fig. 1 a) and
b)) and cesium \cite{Smith} (triangles in Fig. 1 a) and b)) are more
complex, and are less satisfactory reproduced by the
Drude-Sommerfeld dielectric function in the studied frequency range,
as illustrated in Fig. 1 (dotted line for $Li$ and dashed line for
$Cs$).

\begin{figure}[htb]
\begin{center}
    \resizebox{0.75\columnwidth}{!}{\includegraphics{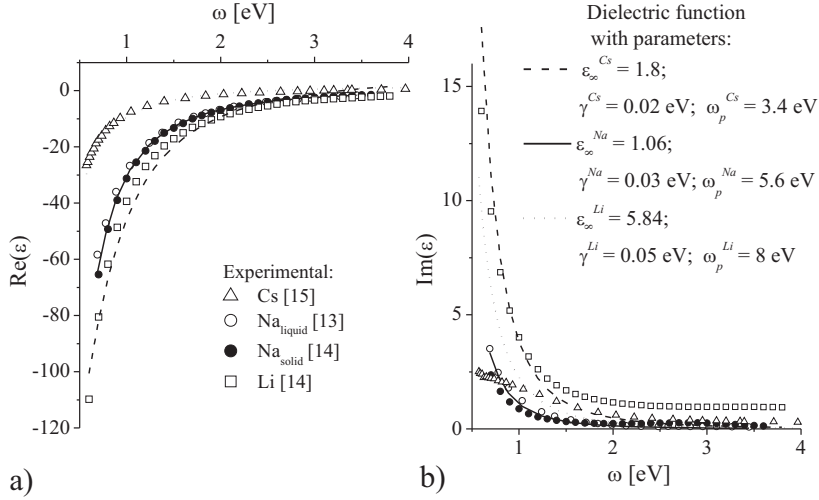} }
\end{center}
\caption{ The real and imaginary part of the dielectric function
with the effective parameters $\varepsilon _{\infty }$, $\gamma $
and $\omega _{p}$ for cesium, sodium and lithium (dashed, solid and
dotted line). Triangles, circles and squares mark Re$(n^{2})$ and
Im$(n^{2})$ values, resulting from measuring the optical constants
$n$ of the corresponding metals [13-15].}
\end{figure}

The proper choice of the parameters entering the dielectric function
is crucial to predicting plasmon resonance characteristics in
experimental realizations. Let's notice, that the optical constants
of metals \cite{InagakiArakawaBirkhoff,InagakiArakawaEmerson,Smith}
were measured in high vacuum conditions and for metals of extremely
clean surfaces. However, optical experiments with metal
nanoparticles are performed usually under less strict laboratory
conditions, for rather contaminated particles. Contamination can be
caused by the presence of the atmosphere
\cite{DerkachovaKolwas,DemianiukKolwas} and as a result of storing
the particles \cite{SonnishenFrantz} or the bulk metal
\cite{DemianiukKolwas} in some liquids before the experiments.
Therefore, the experimental data concerning the plasmon resonance
position can be shifted in respect of the predictions assuming
"ideal" dielectric properties of a metal. Below we discuss the
trends of expected corrections to the plasmon resonance frequencies
due to the modifications in parameters $\varepsilon _{\infty }$,
$\omega_{p}$ and $\gamma $ entering the metal dielectric function
\ref{DrudeDielectricFunction}. We also demonstrate the importance of
the optical properties of environment in determining the position of
plasmon resonance of given polarity $l$.

\section{Results and discussion}

Figures 2 a)-f) illustrate the multipolar ($l=1,2...6$) plasmon
resonance frequencies $\omega _{l}^{\prime }(R)$ and the
corresponding damping rates $\omega _{l}^{\prime \prime }(R)$,
resulting from solving the dispersion equation \ref{DispEquation}
with respect to the frequency allowed being complex. We have used
the M\"{u}ller method of secants for finding the numerical solutions
of $f(v)=0$ assuming the starting approximated values of the
function parameter $v$ in the vicinity of the exact value which may
be complex (the "root" function of the Mathcad program). For given
$l$, the successive values of $R$ were treated as external
parameters and where changed with step $\Delta R\approx 2$ $nm$ up
to the final value of $R=200$ $nm$. The starting, approximated
values for $\omega _{l}^{\prime }(R)$ entering the root procedure
were found from the range between $\omega _{p}/\sqrt{\varepsilon
_{\infty }+\varepsilon _{out}(l+1)/l}$ to $\omega _{p}/\sqrt{2}$ and
the negative values of $\omega _{l}^{\prime \prime }$ were assumed.

\begin{figure}[htb]
\begin{center}
\resizebox{0.75\columnwidth}{!}{\includegraphics{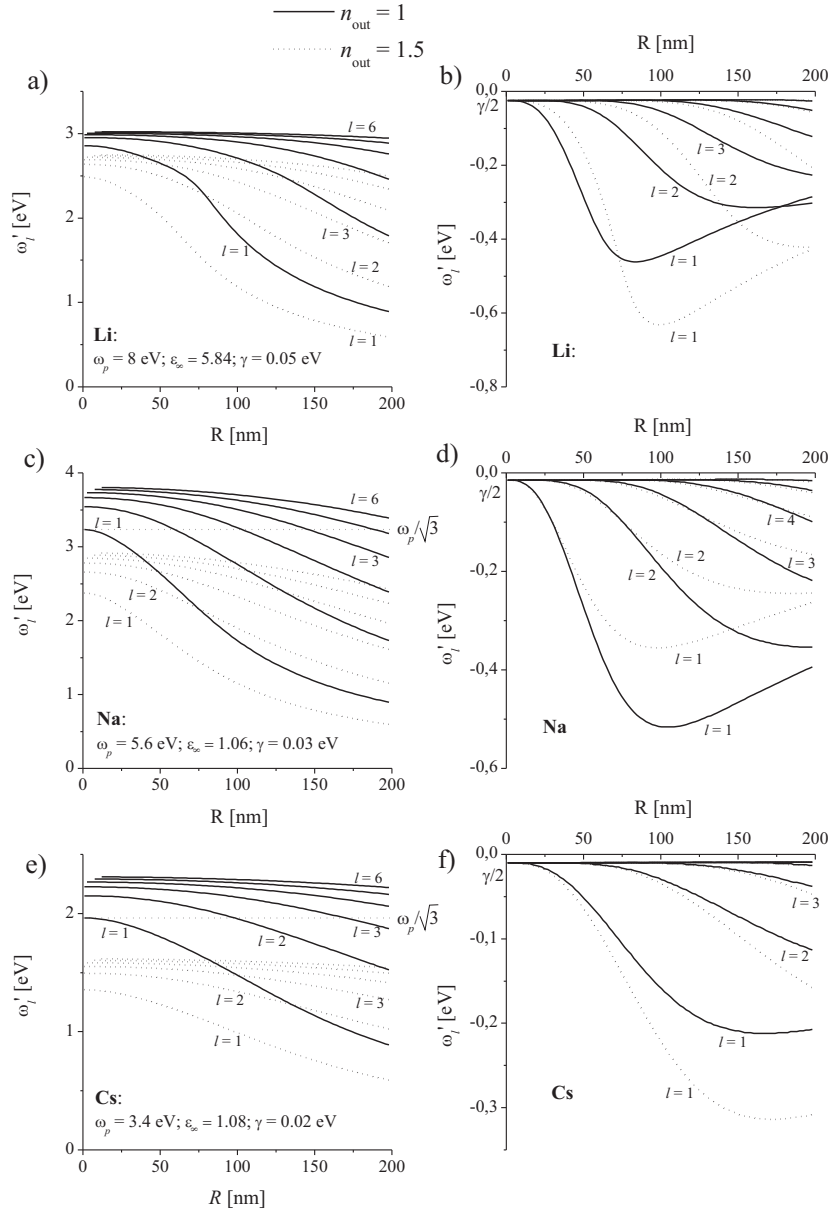}}
\end{center}
\caption{Multipolar plasmon resonance frequencies $\omega
_{l}^{\prime }(R)$ and plasmon oscillation damping rates $\omega
_{l}^{\prime \prime }(R)$ as a function of the radius $R$ (rigorous
modelling) for lithium, sodium and cesium nanospheres in vacuum
($n_{out}=1,$ solid lines), and embedded in glass ($n_{out}=1.5$,
short-dashed lines). The first six ($l=1,2...6$) multipolar plasmon
characteristics are presented.} 
\end{figure}

Plasmon oscillations are always damped (Fig. 2 b), d) and e)) due to
radiation and the relaxation processes included in the relaxation rate $%
\gamma $. The initial increase of the damping rate for given
oscillation mode $l$, is followed by a decrease of $|\omega
_{l}^{\prime \prime }(R)|$ for sufficiently large particles, as
demonstrated for the dipole plasmon damping rate ($l=1$). The
plasmon damping rate dependence on particle size $\omega
_{l}^{\prime \prime }(R)$\ is dominated by the radiative damping
\cite{DerkachovaKolwas}.

Excitation of plasmon resonance in a sphere of given radius $R$
takes place when a frequency $\omega $ of the external
electromagnetic wave fits the frequency of a plasmon mode of given
multipolarity $l$: $\omega =\omega _{l}^{\prime }(R)$. For all
studied simple-metal spheres (Figures 2 a)-f)), plasmon oscillations
can be excited at optical frequencies. For the source of light of
broad spectrum (as in experiments using dark-field microscopic
techniques reported e.g in \cite{SonnishenFrantz} or
\cite{SilveShape}), not only dipole, but also higher multipolar
plasmon resonances can be excited. As we have demonstrated for
sodium spheres in \cite{KolwasDerkachova2}, the highest possible
plasmon multipolar resonance frequency $\omega _{0,l}^{\prime }\ $
and the corresponding damping rate $\omega _{0,l}^{\prime \prime }$
can be attributed to a sphere of a minimum radius $R_{\min ,l}$:
that is: $\omega _{0,l}^{\prime }=\omega _{0,l}^{\prime }(R_{\min
,l})$, $\omega _{0,l}^{\prime \prime }=\omega _{0,l}^{\prime \prime
}(R_{\min ,l})$. $R_{\min ,l}$ being the fast increasing function of
the plasmon multipolarity $l$.

For a given particle size, the frequency of plasmon oscillation
increases with increasing plasma frequency $\omega _{p}$
(free-electron concentration $N$). For example, the dipole plasmon
resonance frequency of a particle of $50nm$ radius is smaller for
cesium than for sodium, both metals with parameter $\varepsilon
_{\infty }$ only slightly differing from 1. With decreasing size the
dipole plasmon oscillation frequencies are slightly modified with
respect to the frequency $\omega _{0,l=1}^{\prime }=\omega
_{p}/\sqrt{3}$ of so called "Mie resonance" \cite{KreibigVollmer}.
However for lithium, with large value of $\varepsilon _{\infty }$,
the dipole plasmon frequency $\omega _{0,l=1}^{\prime }$ is strongly
red shifted with respect to the $\omega _{p}/\sqrt{3}$, as for all
the higher order plasmon frequency dependence upon size.

As demonstrated in Fig. 2 a)-f) (short-dashed lines), the dielectric
properties of the sphere environment can introduce drastic changes
to the multipolar plasmon resonance frequency dependence $\omega
_{l}^{\prime }(R)$ as well as to the corresponding plasmon damping
rates $\omega _{l}^{\prime \prime }(R)$; the proper choice of the
refractive index of the environment is the most effective tool (and
the easiest in practical application) for controlling plasmon
resonance futures.

\begin{figure}[htb]
\begin{center}
\resizebox{0.75\columnwidth}{!}{\includegraphics{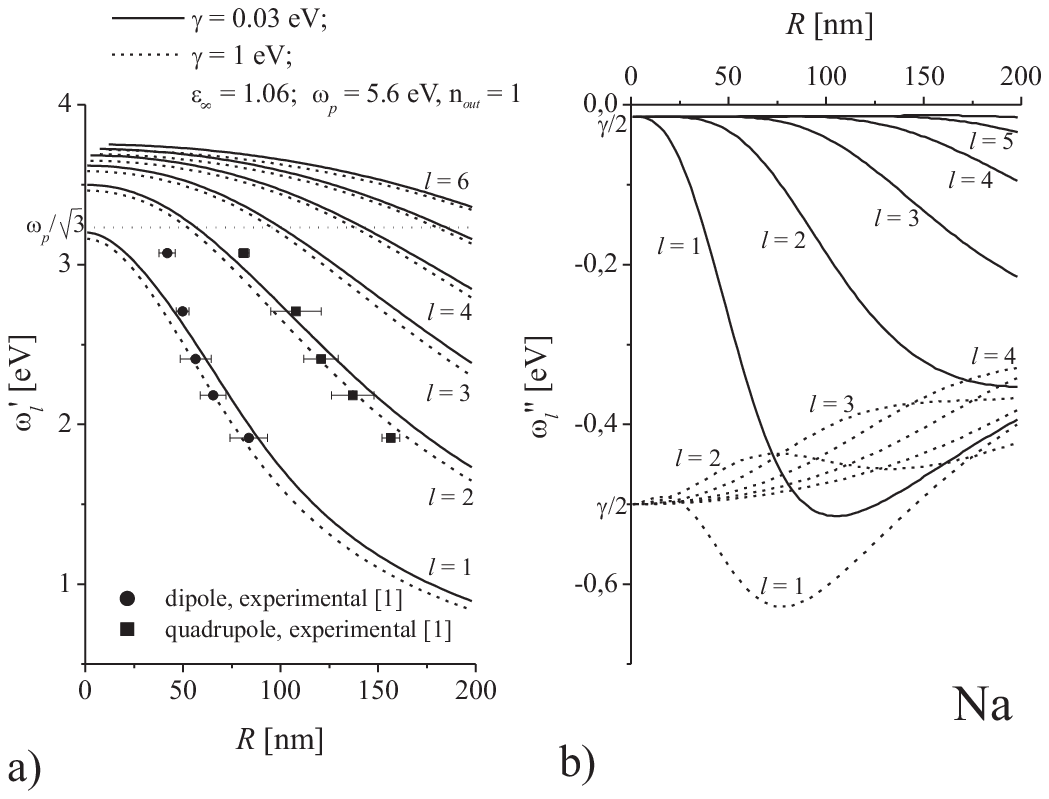}}
\end{center}
\caption{Multipolar plasmon resonance frequencies $\omega
_{l}^{\prime }(R)$ and plasmon oscillation damping rates $\omega
_{l}^{\prime \prime }(R)$ as a function of radius $R$ for very clean
($\gamma =0.03eV$, solid lines) and contaminated ($\gamma =1eV$,
short-dashed lines) sodium nanosphere, calculated for ($l=1,2...6$).
Circles and squares correspond to the sphere radii allowing to
excite the dipole and the quadrupole plasmon resonance with laser
light of different wavelength, according to [1].} 
\end{figure}

As we mentioned above, the experimental results concerning the
multipolar plasmon resonance frequencies for a particle of a given
size can differ from solutions of the eigenproblem with "ideal"
dielectric properties assumed. Our experiment on sodium droplets
which spontaneously grow after the sodium vapour supersaturation by
laser light \cite{DerkachovaKolwas,DemianiukKolwas} can serve as an
example. Due to the presence of the atmosphere and sodium
reactivity, relaxation
rate is increased to the value of $\gamma =1eV$ %
\cite{DerkachovaKolwas,DemianiukKolwas}. It red shifts the plasmon
resonance frequencies $\omega _{l}^{\prime }(R)$ and introduces
important modification to the plasmon damping rates
$\omega_{l}^{\prime\prime }(R)$, as illustrated in Fig. 3 a) and b).
However, if $\gamma \ll \omega $, electron relaxation causes
negligible shift of plasmon resonance frequency, while plasmon
damping rates remain dominated by the size dependence of the
radiative damping, as demonstrated in \cite{KolwasDerkachova2} for
sodium spheres after assuming $\gamma =0$ in the analysis.

The utility value of the elaborated numerical tool for predicting
the multipolar plasmon resonance characteristics depends on the
quality of reproducing the actual optical properties of a metal by
the dielectric function with the effective parameters. It is worth
noting however that such fitting can be reduced to the frequency
range of interest in a particular plasmon application which
corresponds to $1eV\div 4eV$, as illustrated in Fig.2 a), c) and e)
for studied metals.

The elaborated numerical algorithm allows predicting the dependence
of plasmon characteristics upon size of any metal spherical particle
of known form of the dielectric function $\varepsilon (\omega )$.
Such tool can help in tailoring multipolar plasmon resonance
properties according to the requirements of particular application
by choosing the proper size and the material properties of a
nanosphere, as well as the appropriate particle environment.

\end{document}